\documentstyle[preprint,aps]{revtex}
\begin{document} \begin{titlepage}
\begin{flushright}
CU-TP 638
\end{flushright}
\vspace{.5in}

\begin{center}
\Large {\bf Open Charm as a Probe of Pre-Equilibrium Dynamics
in Nuclear Collisions ?}
\end{center}

\begin{center}
{Ziwei Lin and Miklos Gyulassy}\\[2ex]
{Department of Physics, Columbia University, New York, NY, 10027}\\[2ex]
December 15, 1994\\[2ex]
\end{center}

\begin{abstract}
The pre-equilibrium contribution to open charm production in nuclear collisions
at $\sqrt{s}=200$  AGeV is calculated using three different models for the
correlations between momentum and space-time coordinates. Ideal (Bjorken)
correlation between the rapidity $y$  and space-time rapidity $\eta$ of
mini-jet gluons suppresses greatly the pre-equilibrium yield and even allowing
for the minimal uncertainty correlations leads, in contrast to previous
estimates, only to a small pre-equilibrium charm yield as compared to initial
yield due to gluon fusion.  The  ``intrinsic'' charm process is negligible in
the mid-rapidity domain.
\end{abstract}

\end{titlepage}

\section{Introduction}

Open charm production,  direct photon, and dilepton production are among the
most direct probes\cite{shuryak,MW,SX,Geiger}  of the early time evolution of
the quark-gluon plasma  produced in ultra-relativistic nuclear reactions. At
collider energies $\sqrt{s}>200$ AGeV the initial mini-jet  plasma is mostly
gluonic\cite{hotglue,eskola} with a quark content far below its chemical
equilibrium value. Furthermore, the initial transverse momentum distribution of
those gluons is very broad\cite{MW} resembling a  hot thermal gas of gluons
with an effective  temperature $T\sim 500$ MeV \cite{hotglue}. Because charm is
produced mainly through gluon fusion, open charm production provides a probe
of that initial gluonic state. In contrast,  hidden charm\cite{satz} is mostly
sensitive to final state interactions in the later stages of evolution. Photons
and dileptons are complementary probes of  the evolution of the suppressed
quark component of the plasma.

The present study is motivated by two recent studies \cite{MW,Geiger} of open
charm which predicted widely different rates in nuclear collisions. In
ref.\cite{MW} the pre-equilibrium contribution was found to be almost equal to
initial gluon fusion rate.  A similar factor of 2 enhancement of charm from
thermal production in hot-glue scenario was also suggested\cite{hotglue}.  In
ref.\cite{Geiger}, a more provocative claim was made that open charm may
even be enhanced by over an order of magnitude above the initial pQCD rate.
The main result of our present study is that  correlations between the rapidity
$y$ and the space-time rapidity $\eta$ lead to  a large suppression (about a
factor of 40) relative to the uncorrelated case.  Thus,  the pre-equilibrium
open charm production is found to be unfortunately a very small fraction of the
initial fusion rate.  The large enhancement of charm production in
ref. \cite{Geiger} is found to be due to an overestimation of the contribution
from the flavor excitation processes and the use of a low energy $A^\alpha$
scaling from $pp$ reactions measured at $E_{lab}=300-400$ GeV.

The paper is organized as follows: In section~\ref{sec-initial} the dependence
of the direct pQCD rates for charm production on structure functions, $Q^2$
scale, and $K$ factor is reviewed and compared to existing data. The beam
energy dependence and the $A$ dependence of the initial charm production are
compared to results in ref.\cite{Geiger}. In section~\ref{sec-pre}, the
pre-equilibrium charm production is calculated. The mini-jet rapidity and
transverse momentum distribution are fit to results of the Monte Carlo HIJING
model\cite{HIJING} including initial and final state radiation. Three
different models for the space-time and momentum correlations are studied and
the influence on the charm yield is calcualted.  Of the three models, we
concentrate on a minimally correlated model resulting from uncertainty
principle, which is similar to the type of correlation assumed in
ref.\cite{Geiger2}.  We also study the sensitivity of the results to different
models of the formation physics \cite{GyuWang}.  Section~\ref{sec-summary}
contains the summary.

\section {Initial Charm Production}
\label{sec-initial}

Heavy quark production in $pp$ reactions was calculated long ago in
ref.\cite{Combridge} including both fusion and heavy flavor excitation
processes  in the leading order pQCD. It was proposed that the flavor
excitation processes were dominant at high energies because a small $Q^2$
exchange can easily liberate any charm component in the nucleon while gluon
fusion was suppressed because $Q^2 \geq 4M_c^2$.  In the Parton Cascade
Model\cite{Geiger}, both mechanisms are incorporated to calculate $s,c,b$ quark
production in nuclear collisions.  There the results suggested that the flavor
excitation of the charm quark of nuclear structure functions would be the
dominant source of charm production in nuclear collisions as well. However, it
is pointed out\cite{Collins} that the original flavor excitation rates in
ref.\cite{Combridge} were too high in the $x_f\sim 0$ region due to neglected
interference with other pQCD amplitudes to the same order.  When all diagrams
were added together, a large destructive interference was found to suppress the
flavor excitation rates by powers of $\Lambda/M_q$, where $\Lambda \sim 300MeV$
is a typical QCD scale and $M_q$ is the heavy quark mass.  The suppression
factor appears to the process $g + c(\bar c)$ where the charm is evolved from
the structure functions using {\em perturbative} QCD, as also shown in
ref.\cite{Ellis}.  We note that there is a possible non-perturbative charm
component (intrinsic charm) in the nucleon.  There are experimental constraints
on the amount of that non-perturbative charm component\cite{Hoffmann,Ingelman}.
The total contribution of the intrinsic charm was shown in
refs.\cite{Vogt,Brodsky} to be small (about 10\%) in the midrapity region where
most of the charm is made. Although the contribution of the intrinsic charm
component appears important at large $x_f$, its contribution to the total cross
section is small and well within the uncertainties from other sources.

In this paper we only include fusion processes for the parton level cross
sections as in ref.\cite{MW}. For the production in $p-p$ collisions, we use
the light quark and gluon structure functions from Gl\"uck et al.\cite{GRV} and
Duke-Owens\cite{DO} for comparison. The pQCD differential cross sections for
$a+b \rightarrow c\bar c + X$, are taken from ref.\cite{Combridge}.
For example,
\begin{eqnarray}
\sigma_{q\bar q \rightarrow c\bar c}
=\frac{8 \pi \alpha _s^2(Q^2) }{27 \hat s^2} (\hat s + 2 M_c^2)~\chi
\end{eqnarray}
\begin{eqnarray}
\sigma_{g g \rightarrow c\bar c}
=\frac{\pi \alpha _s^2(Q^2)}{3 \hat s} \left[ - (7 + \frac{31 M_c^2}{\hat s})
\frac{1}{4} \chi + (1 + \frac{4 M_c^2}{\hat s} + \frac{M_c^4}{\hat s^2})
\log{\frac{1 + \chi}{1 - \chi}} \right]
\label{EQ:ggfusion}
\end{eqnarray}
where $\chi=\sqrt{1 - 4 M_c^2/\hat s} $ and we consider the following two
choices for the scale $Q^2$ in the coupling constant $\alpha_s(Q^2)=12 \pi/
\left[ (33 - 2 n_f) \log(Q^2/\Lambda ^2) \right]$ from ref.\cite{Combridge}:

1. for $g g \rightarrow c\bar c, Q^2 = \hat s/2$;  for $q\bar q \rightarrow
   c\bar c, Q^2 = \hat s$.  ($Q^2$ choice-1)

2. for both $g g \rightarrow c\bar c$ \& $q\bar q \rightarrow c\bar c, Q^2 =
   \hat s$.  ($Q^2$ choice-2)

We take  $n_f=4$ for charm quark production and $n_f=5$ for bottom quark
production. The QCD scale $\Lambda$ depends on the choice of parton
distribution functions and is given in the table below

\begin{tabular}{|l|c|} \hline
Parton distribution functions    & $\Lambda$(GeV) \\ \hline
GRV-LO set                   & 0.25 \\
GRV-HO set                   & 0.20 \\
Duke-Owens set 1(DO1)            & 0.20 \\
Duke-Owens set 2(DO2)            & 0.40 \\ \hline
\end{tabular}

To incorporate approximately the next-to-leading-order corrections to the above
rates we multiply the leading order results by a K-factor. In general, K-factor
depends on the choice of parton distribution functions, the center of mass
energy of the collision, and the type of the projectile and target particles.
Calculations to order $O (\alpha_s^3)$ for the subprocesses were carried
out\cite{Nason,Beenakker}, and afterwards the calculations to order $O
(\alpha_s^3)$ for $p+p$ collisions were made\cite{Berger,BM}.  For DO1,
$M_c=1.5 $ GeV, $Q^2=4M_c^2, P_{lab} = 100 - 1000$ GeV, the K-factor for $p-p$
collisions \cite{Berger} was found to range from $2.85$ to $4.1$.  We also note
a recent result \cite{SV} where the dependence of the K-factor on the final
momentum of the initially produced charm was studied for high energy $AA$
collisions.  As a function of the rapidity of the charm, the K-factor is almost
a constant $\sim 2$.  As a function of $p_\perp$, the K-factor increases from
$1.3$ at $p_\perp=0.7$ GeV to $3.4$ at $p_\perp=6$ GeV.

In Fig.~1 we compare the so calculated charm cross section to the limited
data on  inclusive $c\bar c$ production  in $p-p$ collisions. The NA34 data for
$\sigma _{charm}$ is taken directly from ref.\cite{NA34}.  The values for the
other data lines are computed from D-meson cross sections according the
argument in ref.\cite{Goshaw} by using the published experiment
results\cite{NA27,E743,E653,E789}.  Earlier experiment results \cite{Tavernier}
also show big uncertainties among the different experiments.

In Fig.~1 We see that both the solid curve and the dashed curve fit the low
energy data reasonably well, so we use these two parametrizations for the
following high energy calculations in this section.  As a consistency check, we
also plot the long-dashed curve using the same parameters as in Fig.~1 of
ref.\cite{Berger} (i.e. DO1, $M_c=1.5 $GeV, $Q^2=4 M_c^2$) using constant $K=3$
for simplicity.  Comparing the solid and dot-dashed curves shows the strong
dependence on the assumed charm quark mass for the GRV-HO set. Comparing the
solid and  dashed curve we see that different choices for the $Q^2$ scale can
be compensated for by shifts in the $K$ factor. These results together with the
large uncertainty of data emphasize the need to measure $pp$ and $pA$ to fix
uncertainties in the initial charm production rate in order that charm
production in $AA$ can be properly calculated.

Next we compare our  results for the rapidity density of produced $c\bar{c}$
pairs at $Y=0$ with results of ref.\cite{Geiger}. In Fig.~2 the energy
dependence in the range between RHIC and LHC (~$\sqrt s = 200 - 6300$ AGeV)
for $Au+Au$ collisions are shown. The scaling from $pp$ results to $AA$ is
\begin{eqnarray}
{\left( \frac{dN}{dY} \right)}^{AA}_{Y=0} = A^{\alpha +1/3}
{\left( \frac{d\sigma}{dY} \right)}^{pp}_{Y=0}/\sigma^{pp}_{inelastic}
\end{eqnarray}
where $\sigma^{pp}_{inelastic}$ is taken from ref.\cite{Goulianos}. Glauber
geometry for central high $A+A$ collisions gives $\alpha=1.$.  In Fig.~2
the solid curve is our result using the same parameters as for the solid curve
in Fig.~1.  The parametrization for the dashed curve in Fig.~1 gives a curve
higher than the solid curve by 15\% to 30\%.  The four long-dashed curves,
curve1 to curve4, are all from PCM calculations\cite{Geiger}. The top curve4 is
the parton cascade model result for the so-called QGP formation case, including
both the fusion and the flavor excitation processes.  That curve is higher than
our solid curve by about an order of magnitude because it includes the
contribution from flavor excitation processes. Curve3, the curve with filled
squares, shows the contribution to curve4 from fusion processes only (processes
(1) and (2) in the notation of ref.\cite{Geiger}), and curve3 is very close to
our results. The bottom curve1 is the estimate without QGP formation by
extrapolating the parton model $pp$ result to $AA$ using $A^{1.09}$ scaling.
It is lower than our solid curve by a factor of $6$ to $2.5$.  The main source
of this difference is from the A-dependence of $p-A$ cross sections.
Ref.\cite{Geiger} used an $A^\alpha$ scaling with $\alpha=0.76$
\cite{smallalpha} instead of the value $\alpha=1$ we use from Glauber
geometry. We note that the value $\alpha=0.76$ is taken from low energy
experiments, where energy conservation suppresses the contribution from
multiple collisions. At high energies, QCD factorization implies that
$\alpha=1$ for $p-A$ scaling is the appropriate scaling modulo small nuclear
shadowing effects.  To demonstrate this effect from different $A^\alpha$
scaling, we multiply Curve1 by a factor of $A^{1.0}/A^{0.76}=3.55$ and get
Curve2, which is close to our results.  In summary, the factor $\sim 50$
enhancement of charm production suggested in ref.\cite{Geiger}  comparing
curve1 with curve4 for charm production at $RHIC$ is a consequence of the
inclusion of incoherent flavor excitation processes and the extrapolation from
$pp$ to $AA$ via low energy scaling.  Given the coherent suppression of the
flavor excitation processes \cite{Collins} and the high energy scaling under
consideration, it is only sensible to compare curve2 with curve3.  In that case
Fig.~2 leads to the expectation that the pre-equilibrium charm production
should be comparable to the initial fusion rate.  This removes the bulk of the
discrepancy between ref.\cite{MW} and ref.\cite{Geiger}.

As a further check on the parameters we compare charmed hadron $x_f$ results in
Fig.~3 with 400 GeV $p-p$ data\cite{NA27} using the idealized
$\delta$-functionfragmentation function.  The realistic fragmentation function
used in ref.\cite{Vogt} lowers the curves slightly and reveals the true
high-$x_f$ intrinsic charm component.  In Fig.~4 we compare $b\bar{b}$
production.  Here we take $M_c=4.75$ GeV as in ref.\cite{BM}, with $K=3,n_f=5$.
The data point at $\sqrt S =630 $GeV is from ref.\cite{UA1}: $\sigma(p\bar p
\rightarrow b + X)=19.3\pm7(exp.)\pm9(th.) \mu b$, and only the experimental
error is indicated in Fig.~4.  At $\sqrt S =1.8 Tev$, our value is  $41.8 \mu b
\times K = 125 \mu b$. This is significantly larger than found in
ref.\cite{BM}.

\section{Pre-equilibrium Charm Production}
\label{sec-pre}

We consider next the pre-equilibrium contribution to the charm yield in $A+A$.
This is the charm produced through final state interactions between partons in
the dense mini-jet plasma.  Here we only calculate the dominant contribution
from mini-jet gluon fusion.

\subsection {Spectrum of Mini-Jets}

The spectrum of mini-jet gluons in leading-order follows from
ref.\cite{CKR}
\begin{eqnarray}
\frac{d \hat \sigma}{d \hat t}_{g g\rightarrow g g}
=\frac{9}{2} \frac{\pi \alpha_s^2}{\hat s^2} \left[3-\frac{\hat u\hat
t}{\hat s^2}-\frac{\hat u\hat s}{\hat t^2}-\frac{\hat s\hat t}{\hat u^2}
\right]
\end{eqnarray}
\begin{eqnarray}
\frac{d \hat \sigma}{d \hat t}_{g q\rightarrow g q}
=\frac{\pi \alpha_s^2}{\hat s^2} \left[ -\frac{4}{9}\frac{\hat u^2+\hat
s^2}{\hat u \hat s}+\frac{\hat u^2+\hat s^2}{\hat t^2} \right]
\end{eqnarray}
The term mini-jets refers to unresolved jets at a scale
$p_\perp>{p_\perp}_{cut}=2$GeV. The inclusive cross section to produce
mini-jets is given by
\begin{eqnarray}
\frac{d \sigma}{d y d p_\perp^2}
= \int d y_3 x_1 f_1 x_2 f_2 \frac{d \hat \sigma}{d \hat t}
(1+2\rightarrow 3+4)
\end{eqnarray}
where $f_1$ is the incident parton distribution evaluated at $x_1=p_\perp(e^y+
e^{y_3})/\sqrt s$ at a scale $Q^2={p_\perp}^2$.  The light-cone coordinates of
the initial and final partons are $p_1=\left[ 2x_1 p_0, 0, \vec 0
\right],~~p_2=\left[ 0, 2x_2 p_0, \vec 0 \right],\;$ $p_3=\left[ m_\perp
e^{y_3},m_\perp e^{-y_3},-\vec {p_\perp} \right] $, and the observed parton has
$p=\left[ m_\perp e^{y},m_\perp e^{-y},\vec {p_\perp} \right] $.  The
subprocess Mandelstam variables are $\hat{s}=s x_1 x_2$ etc..  For the
calculation of mini-jet gluon fusion process in the following
section~\ref{subsec-correlation}, we choose $Q^2=\hat s$.  As in ref.\cite{MW},
we use DO1 as the proton structure functions and $K=2$, $M_c=1.5$GeV for the
mini-jet production.  Shadowing on Au is taken from ref.\cite{shadowing}. The
resulting transverse momentum distribution of mid-rapidity mini-jet gluons at
$\sqrt s= 200$ AGeV is shown by the open circles in Fig.~5.  We call this
distribution the hard distribution since it has ${p_\perp}_{cut}=2$GeV.  It is
compared to the solid line, which is the output of the Monte Carlo calculation
via the HIJING model\cite{HIJING} that includes initial and final state
radiation.

For convenience we have parameterized the Monte Carlo results as the following:
\begin{eqnarray}
\frac{d N}{d y d {\vec {p_\perp}}}\equiv g(p_\perp)  \rho (y) A^{4/3} =
0.06 e^{-1.25 p_\perp} \cos\!{\left[ \frac{\pi (\frac{y}{3.7})^{1.8}}{2}
 \right]} A^{4/3} ,\;\; {\rm with }\; |y| \leq 3.7
\label{EQ:fit}
\end{eqnarray}
In the following, we call this parameterized distribution the soft+hard
distribution. The soft+hard, hard, and Monte Carlo distributions are very close
to each other in the semi-hard  $p_\perp > 2$ GeV region at $y=0$, as seen in
Fig.~5. However the parametrized distribution falls underneath the Monte Carlo
result in the region $p_\perp< 1$ GeV. We emphasize that the soft component is
strongly model dependent as it requires the furthest extrapolation from the
pQCD hard domain. The Hijing yield in that region is due to initial and final
state radiation. Other contributions in this soft domain from coherent string
are possible\cite{eskola}. While most of the following results  are obtained
with the simple parametrization above, we will check the sensitivity to
variations of the soft component as well.  We also note that at bigger rapidity
the $p_\perp$ spectrum falls more rapidly.  The above parametrization does not
include that property.  However, that property only lowers the high $p_\perp$
tail, and hardly changes the low $p_\perp$ part and the total number of the
pre-equilibrium charm.

\subsection{$\eta$ - $y$ correlations}
\label{subsec-correlation}
\subsubsection {Bjorken correlation}

In  ideal Bjorken dynamics, the space-time rapidity $\eta=1/2
\log[(t+z)/(t-z)]$ and the true momentum rapidity $y=1/2\log[(E+p_z)/(E-p_z)]$
are assumed to be perfectly correlated. This is referred to as the
inside-outside picture and the phase-space distribution function in this case
has the form
\begin{eqnarray}
F(\vec x, \vec p, t)_{Bj}
=\frac{(2\pi)^3}{\tau \pi R_A^2 p_\perp} \frac{d N}{dy d {\vec {p_\perp}}}
\delta(\eta -y)  \Theta(\tau-\tau_i) \Theta(\tau_f-\tau)
\end{eqnarray}
$\tau_i=0.1 fm/c$ is the mini-jet formation time.  $\tau_f\approx 1.7 fm/c$ is
the proper time when the energy density of the pre-equilibrium mini-jets falls
by an order of magnitude to $\sim 2 GeV/fm^3$ due to rapid longitudinal
expansion, and that is when we terminate the pre-equilibrium stage.

The phase space distribution is normalized such that
\begin{eqnarray}
\int \frac{F(\vec x, \vec p,t)_{Bj} d^3x}{(2\pi)^3}
=\frac{d^3N}{d^3p}=\frac{1}{E}\frac{d N}{dy d \vec{p_\perp}}
\end{eqnarray}
In this section we study the pre-equilibrium charm production at
$y=0$\cite{MW}:
\begin{eqnarray}
{\left( E\frac{d^3N}{d^3p} \right)}_{y=0}
& = &\int{d^4 x}\int \frac{1}{32(2\pi)^8}
 \frac{d^3{p_1}d^3{p_2}d^3{p^\prime}}{\omega_1
\omega_2 E^\prime} F(\vec x,\vec p_1,t)~F(\vec x,\vec
p_2,t)~|M|^2~\delta^{\!(\!4\!)}\! \left( \!\sum\!P^{\!\mu} \right)
\label{EQ:edndp}
\end{eqnarray}

Denoting $dN/dyd {\vec {p_\perp}} \equiv g(y,p_\perp)$, and $\vec
{{p_\perp}_1}=(\cos\!\phi_1,\sin\!\phi_1,0){p_\perp}_1$, the ideal $\eta-y$
correlation leads to
\begin{eqnarray}
{\left( E\frac{d^3N}{d^3p} \right)}_{y=0}
& = &\int_{\tau_i}^{\tau_f} \frac{d \tau} {32(2\pi)^{\!2} \tau \pi R_A^2}
\int d\eta d {p_\perp}_1 d {p_\perp}_2 d \phi_1 d \phi_2 \frac{
g(\eta,{p_\perp}_1) g(\eta,{p_\perp}_2) \delta\!(\!\sum \!E\!) |M|^2}
{E^\prime} \nonumber \\
& = &\frac {\ln (\tau_f / \tau_i)} {32(2\pi)^{\!2} \pi R_A^2} \int d\eta
d{p_\perp}_2 d \phi_1 d \phi_2 \frac {g(\eta,{p_\perp}_{1,0})
g(\eta,{p_\perp}_2) |M|^2 } {{p_\perp}_2 \left[ 1-\cos\!(\phi_1-\phi_2) \right]
-(E \cosh\!\eta-p\cos\!\phi_1)}
\label{EQ:edndp.delta}
\end{eqnarray}
In deriving the above, we have used kinematic relations
\begin{eqnarray}
E^\prime&=&({p_\perp}_1+{p_\perp}_2) \cosh\!\eta-E \nonumber \\
\frac{\delta\! \left( \!\sum \!E\!\right)}{E^\prime}
&=&\frac{\delta({p_\perp}_1-{p_\perp}_{1,0})} {{p_\perp}_2
\left[ 1-\cos\!(\phi_1-\phi_2) \right]-(E \cosh\!\eta-p\cos\!\phi_1)}
\nonumber \\
{p_\perp}_{1,0}&=&\frac{{p_\perp}_2 ( E \cosh\!\eta-p \cos\!\phi_2)}
{{p_\perp}_2 \left[ 1-\cos\!(\phi_1-\phi_2) \right]-(E
\cosh\!\eta-p\cos\!\phi_1)}
\end{eqnarray}
Numerical integration of  the above integral in equation~(\ref{EQ:edndp.delta})
leads to the results shown in Fig.~6.  The solid line is the
$p_\perp$-distribution for the initial charm production, from
section~\ref{sec-initial}.  We see that the pre-equilibrium contribution in
this strongly correlated case is totally negligible.  This result is similar to
the thermal charm production contribution calculated in ref.\cite{MW} except
that in our case the curve extends to higher $p_\perp$ because of the broader
initial mini-jet distribution in $p_\perp$.

\subsubsection{Uncorrelated $\eta-y$}

In ref.\cite{MW}, another extreme case, opposite to the ideal Bjorken picture,
was considered.  In that case  the gluon distribution is assumed to be
completely uncorrelated as in an ideal thermal fireball.  This assumption leads
to
\begin{eqnarray}
F(\vec x, \vec p, t)_{Fb}
=\frac {(2\pi)^3}{p} \frac{1}{V} \frac{d N}{dy d \vec{p_\perp}}
\end{eqnarray}
If one assumes a fixed volume $V=\tau_i \pi R_A^2$, then $\int dt \sim
\tau_f-\tau_i $, and
\begin{eqnarray}
\int \frac{d^4x}{V^{\!2}} \sim \frac{1}{\pi R_A^2}
\frac{\tau_f}{\tau_i}~~~~\mbox { as in ref.~\cite{MW}. }
\end{eqnarray}
Then from equation~(\ref{EQ:edndp}), we have
\begin{eqnarray}
{\left( E\frac{d^3N}{d^3p} \right)}_{y=0}
=\frac{I(p_\perp)}{32(2\pi)^{\!2}}\int\!{\frac{d^4 x}{V^{\!2}}}
=\frac{\tau_f/\tau_i}{32(2\pi)^{\!2}\pi R_A^2} I(p_\perp)
\label{EQ:edndp.un}
\end{eqnarray}
where
\begin{eqnarray}
I(p_\perp)&=&\int\! \frac{dy_1 dy_2 d{p_\perp}_2 d\phi_1 d\phi_2}{\cosh y_1
\cosh y_2} \frac{ g(y_1,{p_\perp}_{1,0}) g(y_2,{p_\perp}_2) |M|^2} {{p_\perp}_2
\left[ \cosh\! (y_1-y_2) -\cos\!(\phi_1-\phi_2) \right]-(E \cosh\! y_1 -p
\cos\!\phi_1)} \nonumber\\
\frac {\delta \! \left( \!\sum \!E\! \right)}{E^\prime}
&=&\frac {\delta({p_\perp}_1- {p_\perp}_{1,0})} {{p_\perp}_2 \left[ \cosh\!
(y_1 - y_2) -\cos\!(\phi_1-\phi_2) \right] -(E \cosh\! y_1 -p \cos\!\phi_1)}
\nonumber\\
{p_\perp}_{1,0}
&=&\frac{{p_\perp}_2 (E \cosh\! y_2 -p \cos\!\phi_2)} {{p_\perp}_2 \left[
\cosh\! (y_1 - y_2) -\cos\!(\phi_1-\phi_2) \right] -(E \cosh\! y_1 -p
\cos\!\phi_1)}
\label{EQ:pt1.0}
\end{eqnarray}
For uncorrelated case, the pre-equilibrium charm production is much larger
than the Bjorken-correlation case, and is comparable with the initial charm
yield, as shown in Fig.~7.  This is similar to the result in ref.\cite{MW}
where the pre-equilibrium charm production has almost the same magnitude and
$p_\perp$-shape as the initial charm.

\subsubsection{Minimally-correlated $\eta-y$}

We consider here the simplest source of $\eta-y$ correlations resulting from
the minimal geometrical spread in initial production points required by the
uncertainty principle. This type of correlations are included in the parton
cascade model and discussed in ref.\cite{Geiger2}. The  phase space
distribution function including such minimal correlations has the form
\begin{equation}
F(\vec x, \vec p, t)_{Min}
={\cal N}\int\frac{d N}{dy d {\vec {p_\perp}}}
\frac{\theta(\tau_{max}-\frac{t}{\cosh\! y})}
{1+(\frac{t_f(p)}{\Delta t})^2} \rho_0(\vec {x_0},t_0) \delta(\vec x-
\vec{x_0}-\vec{ v} \Delta t) d^3{x_0} dt_0
 \;\;.
\end{equation}
The integration is over the space-time coordinates $(\vec{x}_0,t_0)$ of the
production points of the gluons.  These points are distributed according to a
normalized density $\rho_0(\vec{x}_0,t_0)$. The delta function arises to take
into account the free streaming of the partons from the production point, with
velocity $\vec{v}=\vec{p}/E$, where $E=p_\perp \cosh y$ and  $p_z=p_\perp \sinh
y$. The theta function defines what we mean by pre-equilibrium. The proper time
when the pre-equilibrium fusion is terminated is $\tau_{max}$, which is
determined below in Fig.~8. The theta function insures that only those gluons
with proper time less than $\tau_{max}$ contribute.

The formation physics is included via the Lorentzian formation
factor\cite{GyuWang}
\begin{equation}
[1+(t_f(p)/\Delta t)^2]^{-1} \;\;,
\label{EQ:lorentzian}
\end{equation}
where $\Delta t=t-t_0$ is the elapsed time, and the formation time is given by
\begin{equation}
t_f(p) \simeq \cosh\!y \frac{0.2 {\rm GeV}}{p_\perp} (fm) \;\;.
\end{equation}
We note that the above formation factor more accurately describes the
interference phenomena suppressing production at early time than the
conventionally assumed factor
\begin{equation}
\theta[\Delta t - t_f(p)]
 \;\; .
\label{EQ:theta}
\end{equation}
In the following we consider both formation functions for comparison to check
for the sensitivity to this formation physics.

We assume that  $\int{\rho_0(\vec {x_0},t_0)}d^3\!x_0 dt_0=1$. In this case the
normalization factor is ${\cal N}=(2 \pi)^3/ E $, so that
\begin{equation}
\lim_{t\rightarrow \infty} \int F(\vec x,
\vec p, t)_{Min} d^3 x/(2 \pi)^3 =d^3 N/d^3 p \;\; .
\end{equation}
As discussed in ref.\cite{Geiger2}, the production points are spread along the
beam axis according to the uncertainty principle by an amount $\delta z\equiv d
\sim \hbar /p_\perp$ since the dominant parton interaction leading to a $y=0$
parton with final $p_\perp$ has an initial longitudinal momentum $x P_0 \sim
p_\perp$. We take as a particular model
\begin{equation}
d=\frac{0.2}{p_\perp}(fm)
\end{equation}
Clearly this is only a rough guess, but it allows us at least to investigate
the sensitivity of the results to a particular $\eta-y$ correlation that
results from this spatial spreading of the production points. We emphasize that
it is precisely the uncertainty of the initial space-time formation physics
that leads us to study the possibility of open charm production as an
experimental probe of that physics.

Given the above assumption we take
\begin{eqnarray}
\rho_0(\vec{x_0},t_0)
=\frac{1} {\pi R_A^2} \delta (t_0) \frac{e^{-z_0^2/(2 d^2)}}{\sqrt {2\pi} d}
\end{eqnarray}
where $d$ is the mean spread for gluons depending on $p_\perp$ from above.
This distribution only spreads out the production points along the beam axis. A
more realistic treatment would also smear out in the time coordinate.

Neglecting  transverse expansion, we obtain finally
\begin{eqnarray}
F(\vec x, \vec p, t)_{Min}
=\frac{(2\pi)^3}{\sqrt {2\pi} \pi R_A^2} \frac{p_\perp}{0.2}
e^{-(z-\tanh\!y\:t)^{\!2} (\frac{p_\perp}{0.2})^{\!2}\!/2}
\;\frac{1}{p} \frac{d N}{dy~d {\vec {p_\perp}}}
\frac{\theta(\tau_{max}-\frac{t}{\cosh\! y})} {1+(\frac{0.2 \cosh\!y}{p_\perp
t})^2}
\label{EQ:Fmin}
\end{eqnarray}
Let $a_1=\tanh\!y_1,~a_2=\tanh\!y_2,
{}~b_1=(\frac{{p_\perp}_{1,0}}{0.2})^2/2, ~b_2=(\frac{{p_\perp}_2}{0.2})^2/2$,
then after integration over $z$, we have the final expression as the following,
while its numerical results are shown in Fig.~9:
\begin{eqnarray}
{\left( E\frac{d^3N}{d^3p} \right)}_{y=0}
&=&\frac{\sqrt {\pi}}{16 (2\pi)^4 R_A^2} \int dy_2 d y_1 d{p_\perp}_2
d \phi_2 d \phi_1 \frac{{p_\perp}_{1,0}
\frac{{p_\perp}_{1,0}}{0.2}\;\frac{{p_\perp}_2}{0.2}\; |M|^2}
{{p_\perp}_2 \cosh\!y_1\cosh\!y_2 (E\cosh\!y_2~-p\cos\!\phi_2)}
\nonumber \\
& &\frac{g(y_1, {p_\perp}_{1,0}) g(y_2, {p_\perp}_2)} {\sqrt {b_1+b_2}}
\int_0^{t_f} dt \frac{e^{\frac{-(a_1-a_2)^2 t^2}{1/b_1+1/b_2}}} { \left[
1+(\frac{0.2 \cosh\!{y_1}} {{p_\perp}_{1,0} t})^2 \right]
\left[ 1+(\frac{0.2 \cosh\!{y_2}}{{p_\perp}_2 t})^2 \right] }
\label{EQ:edndp.min}
\end{eqnarray}
In the above $t_f=\tau_{max} \min\,(\cosh\! {y_1}, \cosh\! {y_2})$, and
${p_\perp}_{1,0}$ is the same as in equation~(\ref{EQ:pt1.0}).  Note that by
using the unit $GeV$ for momentum and unit $fm$ for time, the expression
${\left( E\frac{d^3N}{d^3p} \right)}_{y=0}$ in equations
(\ref{EQ:edndp.delta}), (\ref{EQ:edndp.un}), and (\ref{EQ:edndp.min}) has the
dimension $GeV^{-4}fm^{-2}$, and we need a factor $(\hbar c)^2 \sim (0.2GeV
fm)^2$ to convert it to the dimension $GeV^{-2}$, which we have used in Fig.~6,
Fig.~7,Fig.~9 and Fig.~10.

We also plot the energy density curve at $z=0$ as a function of time in
Fig.~8.  We see that it increases first, and reaches maximum at the time
about $0.1 fm/c$, then the energy density decreases linearly to $\sim 2
GeV/fm^3$ at $\sim 0.9 fm/c$ $(~1.7 fm/c~)$ for hard (soft+hard) distribution.
We choose the above time as the cutoff $\tau_{max}$.

The previous uncorrelated case neglects the finite formation times of the
mini-jets.  In order to see the formation-time effect, we also use
the $\theta$-function form in equation~(\ref{EQ:theta}) instead of the
Lorentzian form in equation~(\ref{EQ:lorentzian}) for the formation-time
effect.  The result from this $\theta$-function is about 10\% higher at
$p_\perp=0$GeV, and 10\% lower at $p_\perp=9$GeV, as shown in Fig.~10.  The
lack of sensitivity to the formation-time physics is due to the relative large
$p_\perp$ for the gluon mini-jets in the charm production process.  There would
be more sensitivity had the production been dominated by low $p_\perp$
components.

We also see that for the soft+hard distribution the soft gluons significantly
increase the pre-equilibrium charm production in both low-$p_\perp$ and
high-$p_\perp$ region, with the largest increase in low-$p_\perp$ region.  It
is interesting to identify where the enhancement comes from.  In Fig.~9, the
curve with diamonds shows the contribution from the fusion of soft gluons both
with $p_\perp<2$ GeV, and the curve with unfilled squares shows the
contribution from the fusion of hard gluons both with $p_\perp>2$ GeV.  These
two curves are both very low compared with the curve calculated from the
soft+hard distribution.  So the enhancement going from hard distribution to
soft+hard distribution mainly comes from the fusion of hard and soft mini-jet
gluons.

We have noted before that our fit for the mini-jet gluon spectrum falls below
the Monte Carlo result from HIJING calculation.  We can fit the soft gluons
from HIJING better by using $0.265 e^{-2.6 p_\perp}$ for $p_\perp \in (0, 1.1)$
GeV, and use the old fit $0.06 e^{-1.25p_\perp}$ for higher-$p_\perp$ gluons.
This new fit gives us more very soft gluons.  We have done the calculation for
minimally-correlated case using the new fit, and the result is different
only by less than 10\%, which means the super-soft gluons are not very
important for the pre-equilibrium charm production.

There is also a possible cross-term contribution from the interactions of the
incoming nuclei and the pre-equilibrium gluon mini-jets.  However, our
preliminary result shows that it is not larger than the above pre-equilibrium
charm yield and is therefore also negligible compared with the initial charm
production.

\subsection{Why is the pre-equilibrium charm yield so small?}

To understand the  reason why the pre-equilibrium charm yield
is so small compared to the initial yield as found through tedious numerical
calculations in the previous section,
 we consider here the calculation of the total
number of  pre-equilibrium charm pairs.
The expression for that number is given by
\begin{eqnarray}
N=\frac{(\hbar c)^2}{4 (2 \pi)^6} \int d^4x \int \frac{d^3 p_1}{\omega_1}
\frac{d^3 p_2}{\omega_2} F(\vec x, \vec p_1, t) F(\vec x, \vec p_2, t) \hat s
\hat \sigma(\hat s)
\label{EQ:nccbar}
\end{eqnarray}
where $\hat \sigma(\hat s)$ is the integrated cross section for the process $gg
\rightarrow c \bar c$, see equation(\ref{EQ:ggfusion}).
Our main strategy  is to estimate the mean difference between the two gluon
rapidities, then from the kinematical
constraint on charm production ($\hat s \geq 4 M_c^2$) estimate
the effective lower cutoff  for $p_\perp$ of the mini-jet gluons.  Thus we
separate the $p_\perp$ integrals from the rapidity integrals.
 and have a rough estimate for the total number of charm pairs.

\subsubsection{Bjorken correlation case and Uncorrelated case}
For the fireball case,
\begin{eqnarray}
F(\vec x, \vec p, t)_{Fb}
=\frac{(2\pi)^3}{p}\frac{1}{V} \frac{d N}{dy d \vec{p_\perp}}  \nonumber\\
\hat s=2 {p_\perp}_1 {p_\perp}_2 \left[ \cosh(y_1-y_2) -
\cos(\phi_1-\phi_2) \right]
\label{EQ:hats}
\end{eqnarray}
For the Bjorken case,
\begin{eqnarray}
F(\vec x, \vec p, t)_{Bj}
=\frac{(2\pi)^3}{p_\perp}\frac{d N}{dy d \vec{p_\perp}} \frac{\delta(\eta
-y)}{\tau \pi R_A^2} \Theta(\tau-\tau_i) \Theta(\tau_f-\tau) \nonumber\\
\hat s=2 {p_\perp}_1 {p_\perp}_2 \left[1 - \cos(\phi_1-\phi_2) \right]
\end{eqnarray}
For all the cases, we use the  fit to the
gluon distribution given by  equation(\ref{EQ:fit}),
where $
g(p_\perp) \equiv a e^{-b p_\perp} = 0.06 e^{-1.25 p_\perp}$.
Therefore,
\begin{eqnarray}
N_{Fb}&=&\frac {(\hbar c)\!^2 \frac{\tau_f}{\tau_i} A^{8/3}}{4 \pi R_A^2}
\int \! dy_1 \frac {\rho (y_1)}{\cosh {y_1}}
\int \! dy_2 \frac {\rho (y_2)}{\cosh {y_2}}
\int \! d{p_\perp}_1 g({p_\perp}_1) \int \! d{p_\perp}_2 g({p_\perp}_2)
\int \! d\phi_1 \int \! d\phi_2 \hat s \hat \sigma (\hat s)
\nonumber\\
N_{Bj}&=&\frac {(\hbar c)\!^2 \ln \!\frac {\tau_f}{\tau_i} A^{8/3}}{4 \pi
R_A^2}  \int \! d \eta \left[ \rho (\eta) \right]^2
\int \! d{p_\perp}_1 g({p_\perp}_1) \int \! d{p_\perp}_2 g({p_\perp}_2)
\int \! d\phi_1 \int \! d\phi_2 \hat s \hat \sigma (\hat s)
\end{eqnarray}
The dominant contribution is coming from the vicinity of the production
threshold where $\hat s=4 M_c^2=9 GeV^2$\cite{Combridge}, so we make the
following rough estimates:
\begin{eqnarray}
\hat s \hat \sigma (\hat s) \sim \alpha ^2(\hat s)  \sim 0.06\; \nonumber\\
\int d{p_\perp} g(p_\perp) \sim \int_{p_c}^\infty d{p_\perp}
g(p_\perp)  \sim \frac {a}{b e^{b p_c}}
\end{eqnarray}
where $p_c$ is the effective cutoff value
for ${p_\perp}_1$ and ${p_\perp}_2$ from the requirement $\hat s \geq 4
M_c^2$.

For the fireball case,
\begin{eqnarray}
\langle {\cosh (y_1-y_2)} \rangle&\equiv &
\frac {\int \! dy_1 dy_2 \frac {\rho (y_1)}{\cosh {y_1}} \frac {\rho
(y_2)}{\cosh
{y_2}} \cosh (y_1-y_2)}{\int \! dy_1 dy_2 \frac {\rho (y_1)}{\cosh {y_1}}
\frac
{\rho (y_2)}{\cosh {y_2}} }
\sim 4.0
\label{EQ:rapidity}
\end{eqnarray}
Since the mini-jet $p_\perp$ spectrum is dropping almost exponentially, the
production heavily favors the smaller cutoff $p_c$, so the mean value of $\cos
(\phi_1-\phi_2)$ is most likely to be negative.  We take $\langle {\cos
(\phi_1-\phi_2)} \rangle \sim -0.5 $.  Then $\hat s \sim 9 p_c^2$, so the
effective cutoff for the fireball case is $p_c \sim 1.0$ GeV.

On the other hand, for the Bjorken case,
\begin{eqnarray}
y_1=y_2=\eta \;\;\Rightarrow \hat s \sim 3 p_c^2
\;\;\Rightarrow p_c \sim 1.73 {\rm GeV}
\end{eqnarray}
Using the same values as in section~\ref{subsec-correlation}: $\tau_i=0.1 fm$,
$\tau_f=1.0 fm$ for fireball case, $\tau_f^{\prime}=1.7 fm$ for Bjorken case,
and
\begin{eqnarray}
\int \! dy_1 \frac {\rho (y_1)}{\cosh {y_1}} \sim 2.8, \;\;
\int \! d \eta \left[ \rho (\eta) \right] ^2 \sim 4.9
\end{eqnarray}
We then have the estimate for the total number of the pre-equilibrium charm:
\begin{eqnarray}
N_{Fb}&\sim&\frac {(\hbar c)\!^2 \frac {\tau_f}{\tau_i} A^{8/3}}{4 \pi R_A^2}
\;2.8^2 (\frac{a}{b e^{1.0b}})^2 (2\pi)^2 \left[ \hat s \hat \sigma (\hat s)
\right] \sim 3.5        \nonumber\\
N_{Bj}&\sim&\frac {(\hbar c)\!^2 \ln \!\frac{\tau_f^{\prime}}{\tau_i} A^{8/3}}
{4 \pi R_A^2} \;4.9 (\frac{a}{b e^{1.73b}})^2 (2\pi)^2 \left[ \hat s \hat
\sigma (\hat s) \right] \sim 0.098
\end{eqnarray}
Therefore we estimate $N_{Fb}/N_{Bj} \sim 35$, in rough agreement with the
detailed numerics.  We see that the main source of the large increase going
from the Bjorken case to the fireball case comes from the different $p_\perp$
cutoff.  In the uncorrelated fireball case, one allows particles with different
rapidities to interact with each other (see equation(\ref{EQ:rapidity})), thus
more low $p_\perp$ gluons can take part in the interaction.  Since the mini-jet
$p_\perp$ spectrum is dropping almost exponentially, the fireball case produces
a lot more pre-equilibrium charm than the Bjorken case (a factor of 6 increase
from the smaller $p_\perp$ cutoff).  Although the questionable linear proper
time dependence in the fireball case also gives an considerable increase (about
a factor of 3.5), it is not as important as the correlation effect.

\subsubsection{Minimal correlation case}
For the minimal correlation case, the estimate is unfortunately not as
straightforward.    The phase space distribution function is
\begin{eqnarray}
F(\vec x, \vec p, t)_{Min}
=\frac{(2\pi)^3}{\sqrt {2\pi} \pi R_A^2} \frac{e^{-(z-\tanh\!y\:t)^{\!2}
(\frac{p_\perp}{\hbar c})^{\!2}\!/2}}{\hbar c \cosh y}
\;\frac{d N}{dy~d {\vec {p_\perp}}} \theta(\tau_{max}-\frac{t}{\cosh\! y})
\theta(\frac{\hbar c \cosh\! y}{p_\perp}-\tau_{max})
\end{eqnarray}
and $\hat s$ is the same as in equation(\ref{EQ:hats}).  In the above
distribution function we choose to use the $\theta$-function for the
formation-time effect.  We have seen from Fig.~10 that Lorentzian
formation-time formula and $\theta$-function formula give almost the same
result.

Using equation(\ref{EQ:nccbar}) and after the integration over $z$, we have
\begin{eqnarray}
N_{Min}=\frac {(\hbar c)^2 A^{8/3}} {4 \pi R_A^2 \sqrt \pi}
&\int& \! d{p_\perp}_1 g({p_\perp}_1) \int \! d{p_\perp}_2
g({p_\perp}_2)
\int \! d\phi_1 \int \! d\phi_2 \nonumber\\
\times &\int& \! dy_1 \frac {\rho (y_1)}{\cosh {y_1}}
\int \! dy_2 \frac {\rho (y_2)}{\cosh {y_2}}  \hat s \hat \sigma (\hat s)
\int_{t_{min}}^{t_{max}} \! dt\; \frac {e^{\frac{-(a_1-a_2)^2
t^2}{1/b_1+1/b_2}}} {\sqrt {1/b_1+1/b_2}}
\label{EQ:NMin}
\end{eqnarray}
where
\begin{eqnarray}
t_{min}=\hbar c \max \left( \frac{\cosh y_1}{{p_\perp}_1},\frac{\cosh
y_2}{{p_\perp}_2} \right), t_{max}=\tau_f \min (\cosh y_1, \cosh y_2)
\end{eqnarray}
and $a_1,a_2,b_1,b_2$ are defined the same as in equation(\ref{EQ:Fmin}).

We estimate that for the dominant part of the integral
\begin{eqnarray}
\frac{1}{b_1}+\frac{1}{b_2} \sim \left( \frac {2 \hbar c}{p_c} \right)^2,
t_{min} \sim \frac{\hbar c \cosh \bar y}{p_c},
t_{max} \sim \tau_f \cosh \bar y
\end{eqnarray}
where $\bar y=(|y_1|+|y_2|)/2$.
Now let $u=t \;p_c/(\hbar c \cosh \bar y)$, then the last 3-dimensional
integral in equation(\ref{EQ:NMin}) without the factor $\hat s \hat \sigma
(\hat s)$ is
\begin{eqnarray}
J \sim
\int \! dy_1 \frac {\rho (y_1)}{\cosh {y_1}}
\int \! dy_2 \frac {\rho (y_2)}{\cosh {y_2}} \frac {1}{2} \cosh \bar y
\int_{1}^{\frac{\tau_f p_c}{\hbar c}} du\; e^{-[ \frac{\sinh (y_1-y_2)
\cosh \bar y }{2 \cosh y_1 \cosh y_2} ] ^2 u^2}
\end{eqnarray}
The $u$-integral gives a dependence on $\tau_f$ which is similar to the
logarithmic dependence in the Bjorken case, and the exponential form in the
integrand forces the spread $y_1-y_2$ to be small. Numerically, by taking
$\tau_f \sim 1.7 fm/c$ , $p_c \sim 2.0$ GeV (as the first-step value) in the
$u$-integral, the above 3-dimensional integral is $J \sim 19.1$ , and when the
integrand is weighed by $\cosh (y_1-y_2)$, the integral is $\sim 23.3$.  So
\begin{eqnarray}
\langle {\cosh (y_1-y_2)} \rangle \sim 23.3/19.1 \sim 1.22
\Rightarrow \hat s &\sim& 3.44 p_c^2  \Rightarrow p_c \sim 1.62 {\rm GeV}
\end{eqnarray}
Note that the above determined value of $p_c$ is insensitive to the
first-step $p_c$ value we tried in the $u$-integral.

Therefore for the total pre-equilibrium charm number,
\begin{eqnarray}
N_{Min} &\sim& \frac {(\hbar c)^2 A^{8/3}} {4 \pi R_A^2 \sqrt \pi}
\int \! d{p_\perp}_1 g({p_\perp}_1) \int \! d{p_\perp}_2
g({p_\perp}_2) \int \! d\phi_1 \int \! d\phi_2
\;J\;\hat s \hat \sigma (\hat s)
\nonumber\\
&\sim& \frac {(\hbar c)^2 A^{8/3}} {4 \pi R_A^2}
\frac {J}{\sqrt \pi} (\frac{a}{b e^{1.62b}})^2 (2\pi)^2 \left[ \hat s \hat
\sigma (\hat s) \right]  \sim 0.10
\end{eqnarray}
Therefore $N_{Min}/N_{Bj} \sim 1$.
{}From the above estimate we can see that although the minimally-correlated
case
allows particles with different rapidities to interact, the dominant
contribution still comes from the region where the two gluons have almost the
same rapidity, thus there is no sizeable enhancement in the pre-equilibrium
charm yield.  The Minimally-Correlated case is very much like the Bjorken case
in that the dominant contribution comes from $y_1 \simeq y_2$ region.

As a comparison to the above rough estimates in this section, the numerical
integration gives $N_{Fb}=3.8$, $N_{Bj}=0.093$, and $N_{Min}=0.078$, so
$N_{Min}/N_{Bj} \sim 80\%$.

\section {Discussion and Summary}\label{sec-summary}

In this paper, we calculated initial and pre-equilibrium charm production in
nuclear collisions to test the sensitivity of this probe to the unknown initial
conditions in such reactions.  For the initial charm production, the sensitive
dependence on the choice of structure functions, the $Q^2$ scale, and the
K-factor was noted. The parameters were fixed by fitting the limited available
experimental data at lower energies.  We emphasized the need for new
measurements of $pp$ and $pA$ charm production to reduce the present large
theoretical uncertainties.  We argued that the copious charm production
predicted in ref.\cite{Geiger} was mainly due to the neglect of the coherent
suppression of  flavor excitation processes.  Our calculated initial charm
yields are close to those computed in ref.\cite{MW} and to the curve 2 in
Fig.~2 from ref.\cite{Geiger}.

For the contribution from pre-equilibrium charm production, we studied the
effect of correlations between the rapidity $y$ and space-time rapidity $\eta$
of mini-jet gluons. For the ideal Bjorken-correlated case, where
$\eta=y_1=y_2$, the pre-equilibrium charm production is negligible compared
with the yield due to initial gluon fusion.  For the opposite extreme fireball
case, corresponding to uncorrelated $y$ and $\eta$,  the pre-equilibrium charm
production is almost a factor of 50 larger than in the Bjorken-correlated case
and is comparable with the initial charm yield\cite{MW}.  By the estimates of
the total pre-equilibrium charm number, we found the the difference mainly
comes from the $\eta-y$ correlation. Therefore, the pre-equilibrium charm
production is very sensitive to the $(\eta-y)$ correlations in the initial
state.

In order to investigate the effect of more realistic correlations
that may exist in the initial mini-jet plasma, we introduced a minimal
correlation model taking into account  the uncertainty principle along the
lines of ref.\cite{Geiger2}. Our main result is that this minimal correlation
is similar to the ideal Bjorken correlation case and produces negligible
pre-equilibrium charm compared with the initial charm yield.  We also found
that the pre-equilibrium charm yield is rather insensitive to the formation
physics because the early-formed $p_\perp > 1$ GeV gluons dominate.

Acknowledgements: We thank K. Geiger, B. M\"uller, X.N. Wang, L. Xiong for
useful discussions and A. Mueller for bringing refs.\cite{Collins,Brodsky} to
our attention.

\pagebreak
%
{}

\pagebreak

\section*{Figure Captions}
\begin{description}
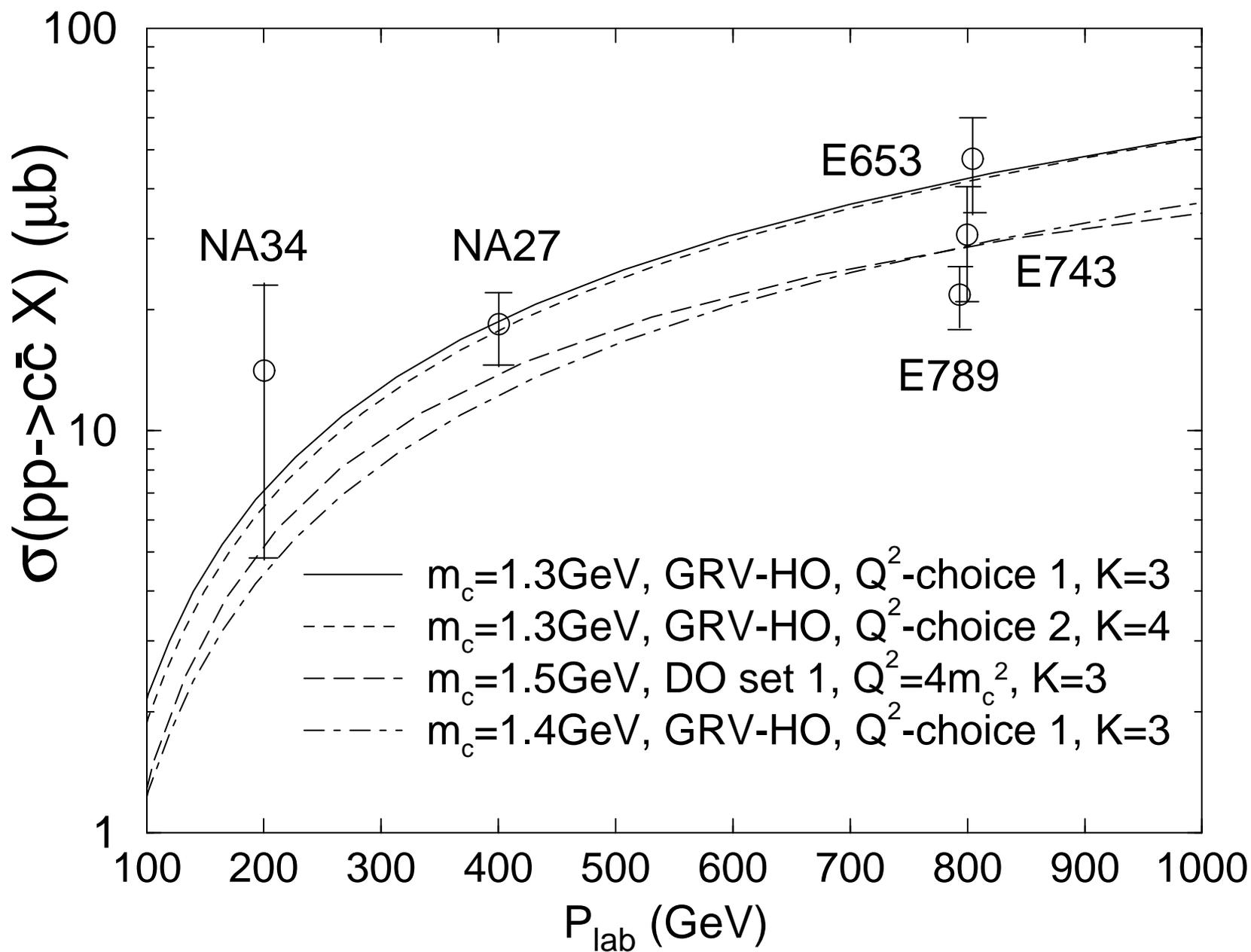
\item[Fig.1]
The cross section for $pp \rightarrow c\bar c X$ is plotted as a function of
$P_{lab}$.  The solid line is our result with $M_c = 1.3$ GeV, $K=3$, $Q^2$
choice-1 and GRV-HO set.  The long-dashed curve is the result with the same
parameters as in Fig.~1 of \cite{Berger}, but using a K-factor of 3 instead of
doing $O(\alpha_s^2)$ calculation.
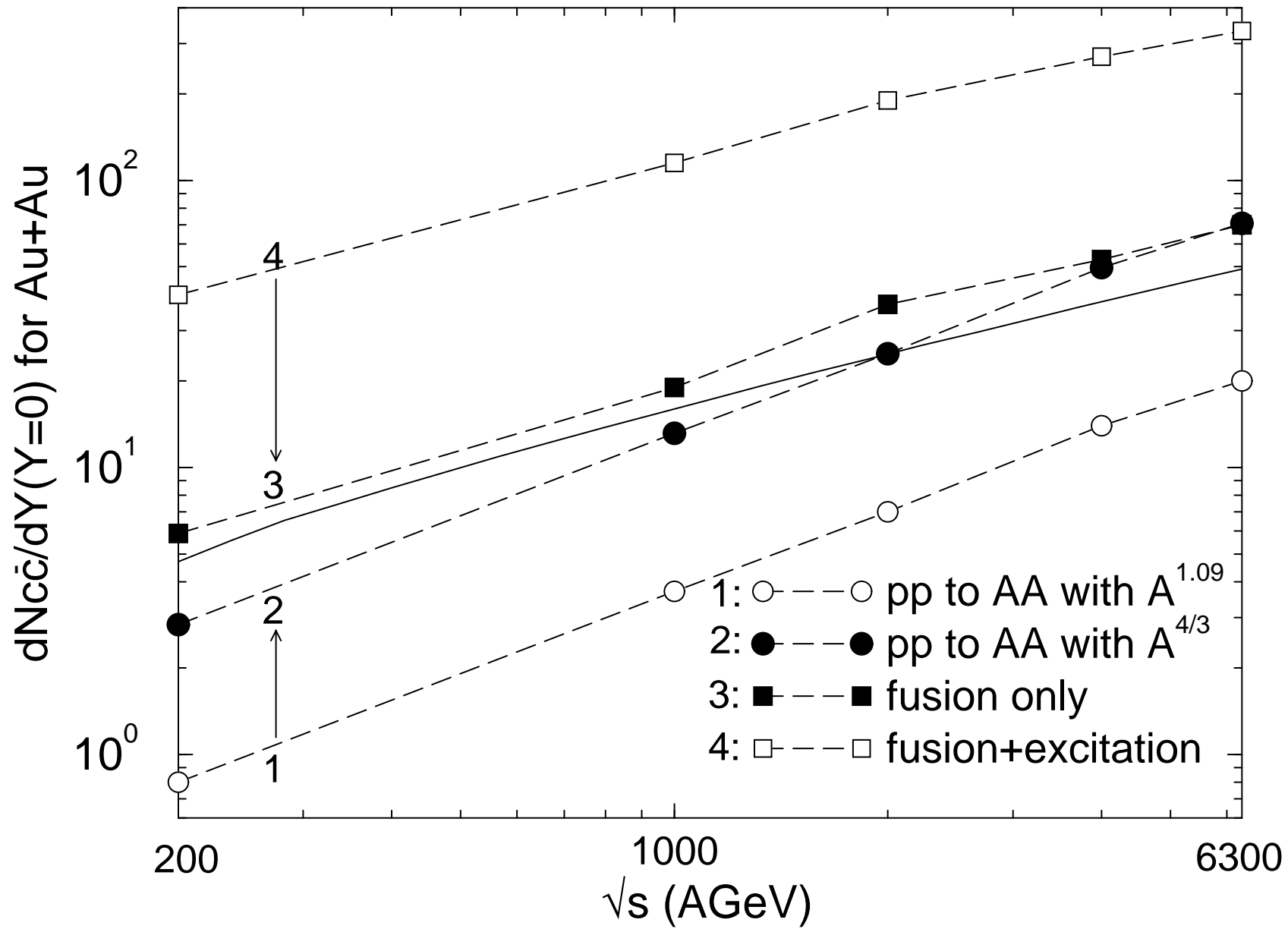
\item[Fig.2]
${\left( dN_{c\bar c}/dY \right)}_{Y=0}$, rapidity density of charm and
anticharm pairs for $Au-Au$ collisions vs $\sqrt S/A$.  Curves 1-4 from the
calculation of Parton Cascade model\cite{Geiger} are compared to our
calculation of the yield (the solid line)due to initial fusion processes.  The
top curve4 is the total charm production with QGP formation including the
incoherent flavor excitation processes.  Curve3 shows the charm production in
the case of QGP formation without excitation processes.  The bottom curve1 is
the parton model result extrapolated to $AA$ from $pp$ using the $A^{0.76}$
scaling measured at much lower energies.  Curve2 is the parton model result
scaled by $A^{4/3}$. Our curve uses the asymptotic $A^{4/3}$ scaling.  As shown
by the two arrows, curve4 becomes curve3 when the coherent cancellation of
flavor excitation processes is considered, and curve1 becomes curve2 when the
high energy scaling is used.  So the net dynamical enhancement in the PCM (by
comparing curve3 to curve2) is comparable to the result of ref.\cite{MW}.
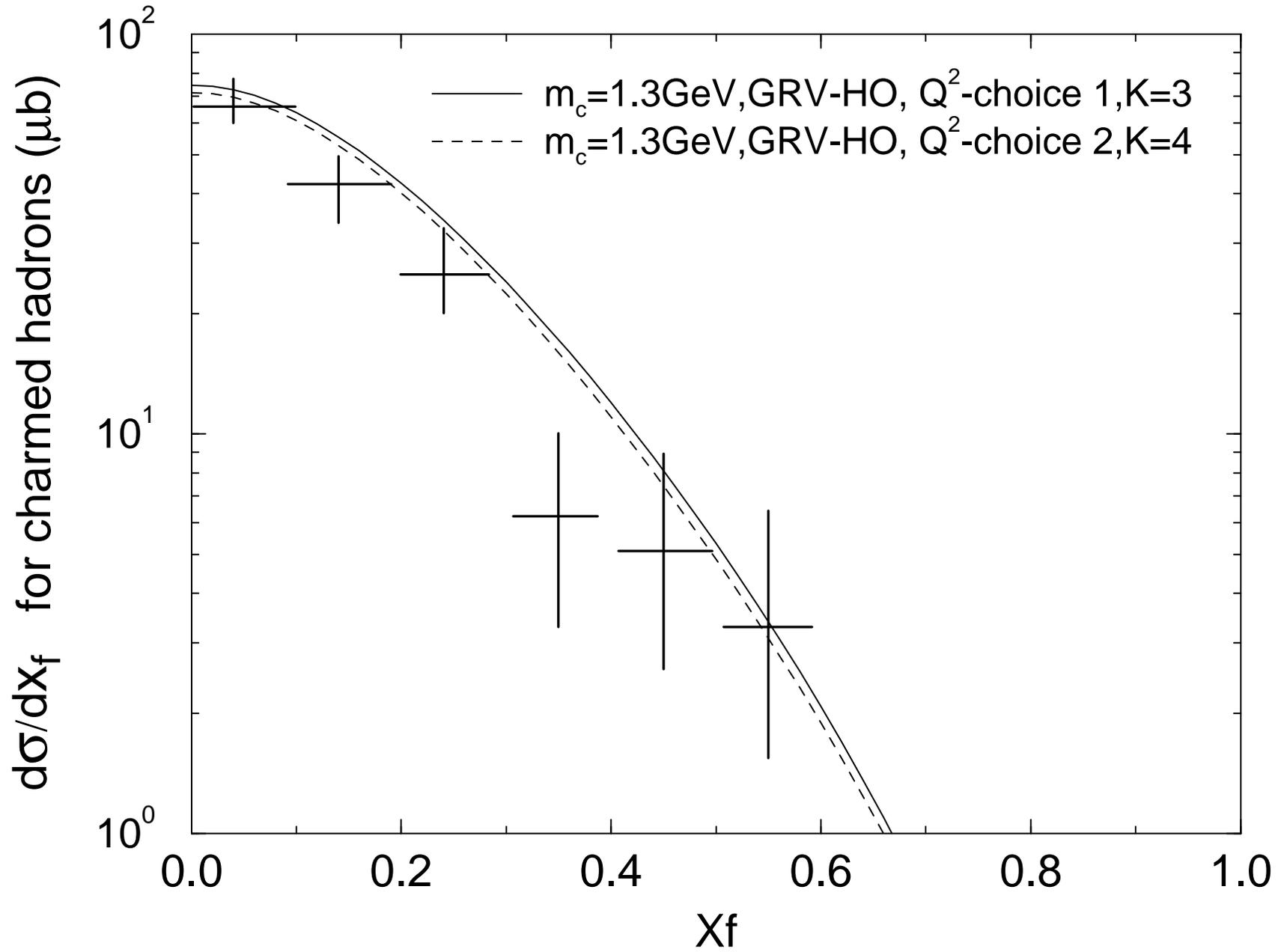
\item[Fig.3]
The production of charmed hadrons as a function of $x_f$ for $p-p$ collisions
at $P_{lab}=400$ GeV\cite{NA27}.  The solid curve is our result for $d\sigma/d
x_f$ using the first parameterization.  The dashed curve is our result using
the second parameterization.  These curves assume a delta function charm
fragmentation function.

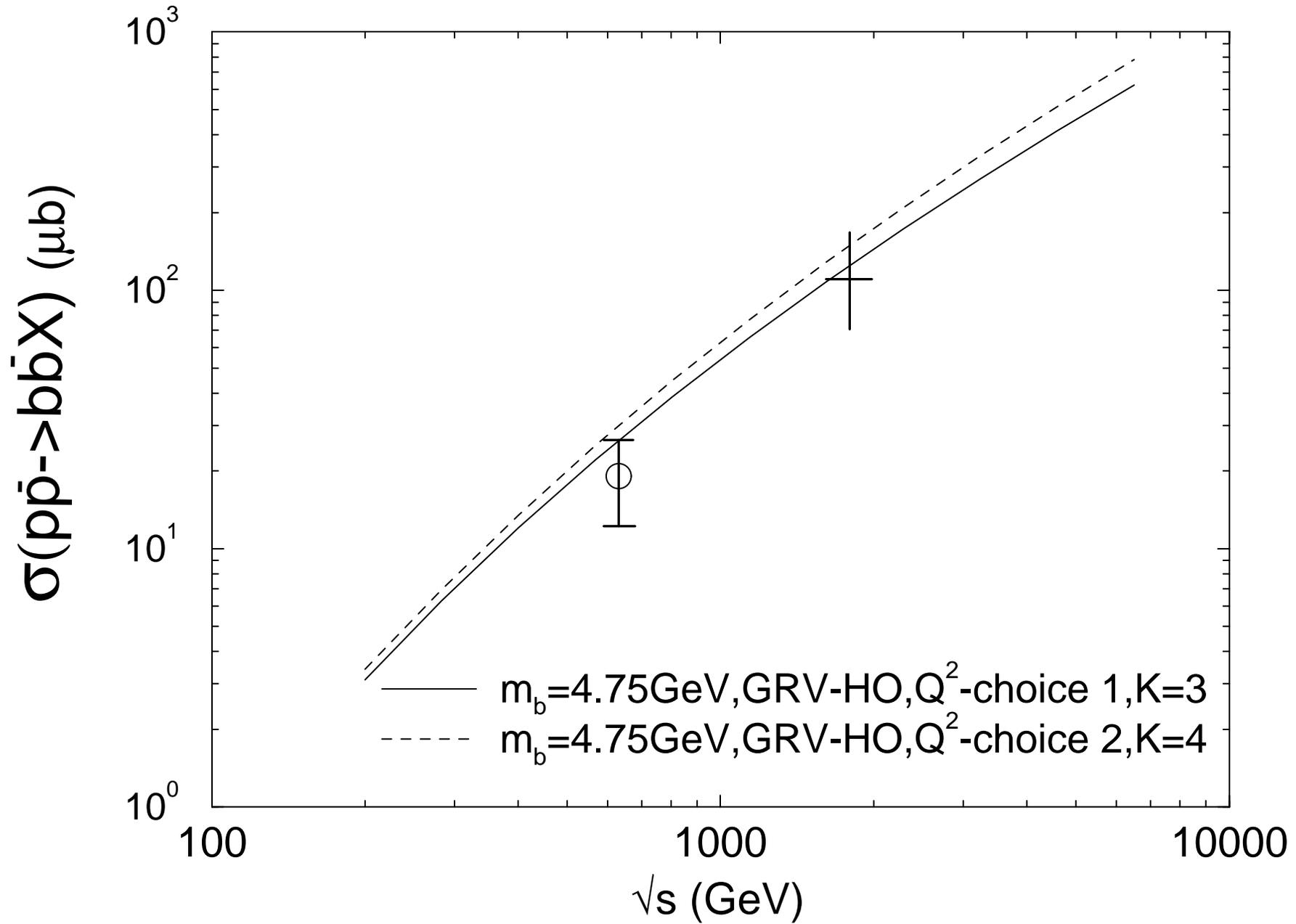
\item[Fig.4]
The cross section for $p\bar p \rightarrow b\bar b+X$ vs $\sqrt S/A$.  The data
point at $\sqrt S=630$ GeV is from ref.\cite{UA1}.  The dashed cross at $\sqrt
S=1.8$ Tev is obtained indirectly from \cite{BM}, and the error bar is only
illustrative.
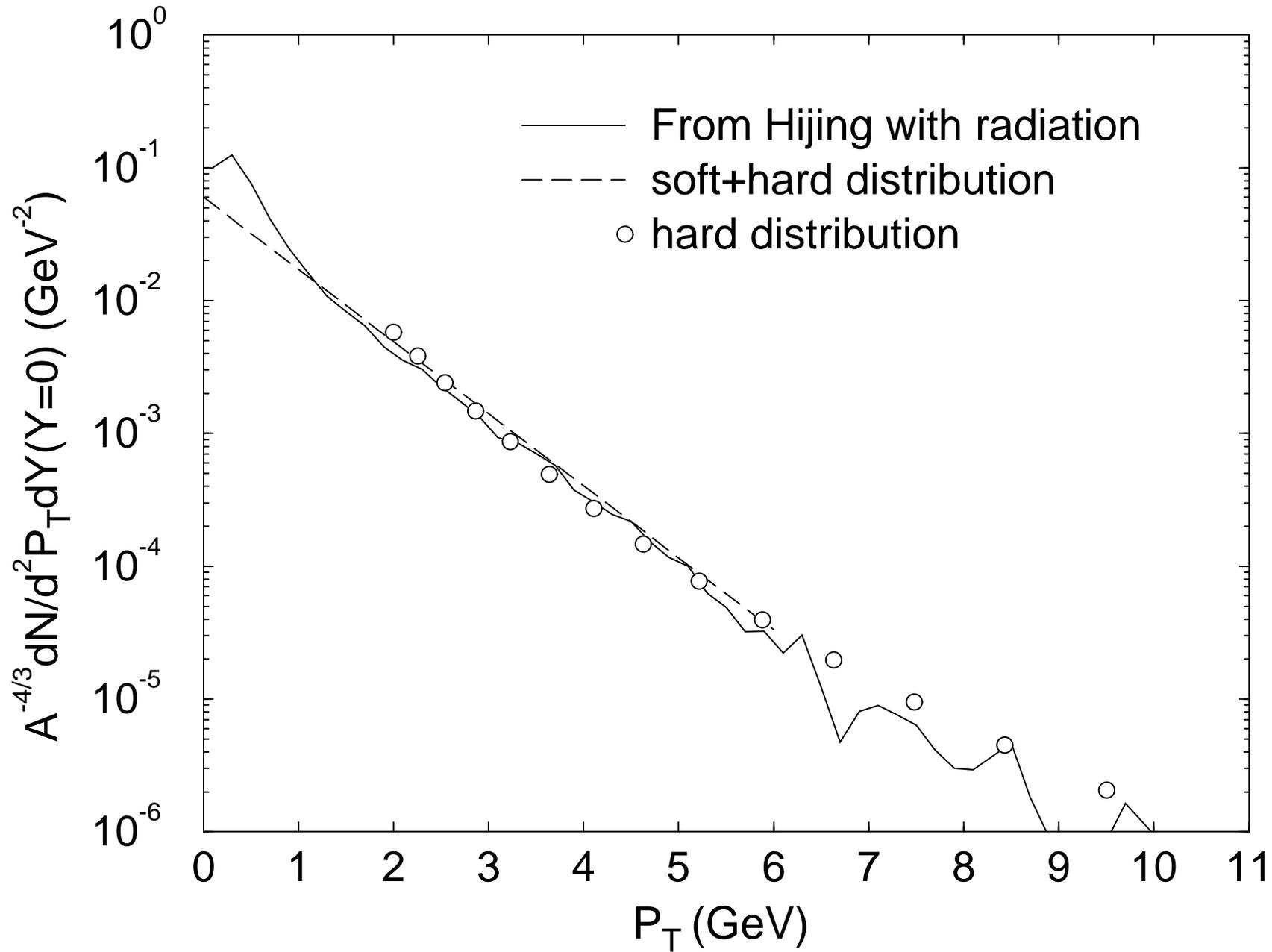
\item[Fig.5]
The mini-jet gluon distribution $A^{-4/3}\left( dN/dy d \vec{p_\perp}
\right)_{y=0}$ is plotted. The solid curve is taken from the HIJING calculation
with radiation effects  included, and the circles are our result from the
initial production. The dashed line is the fit  $0.06 e^{-1.25p_\perp}$.
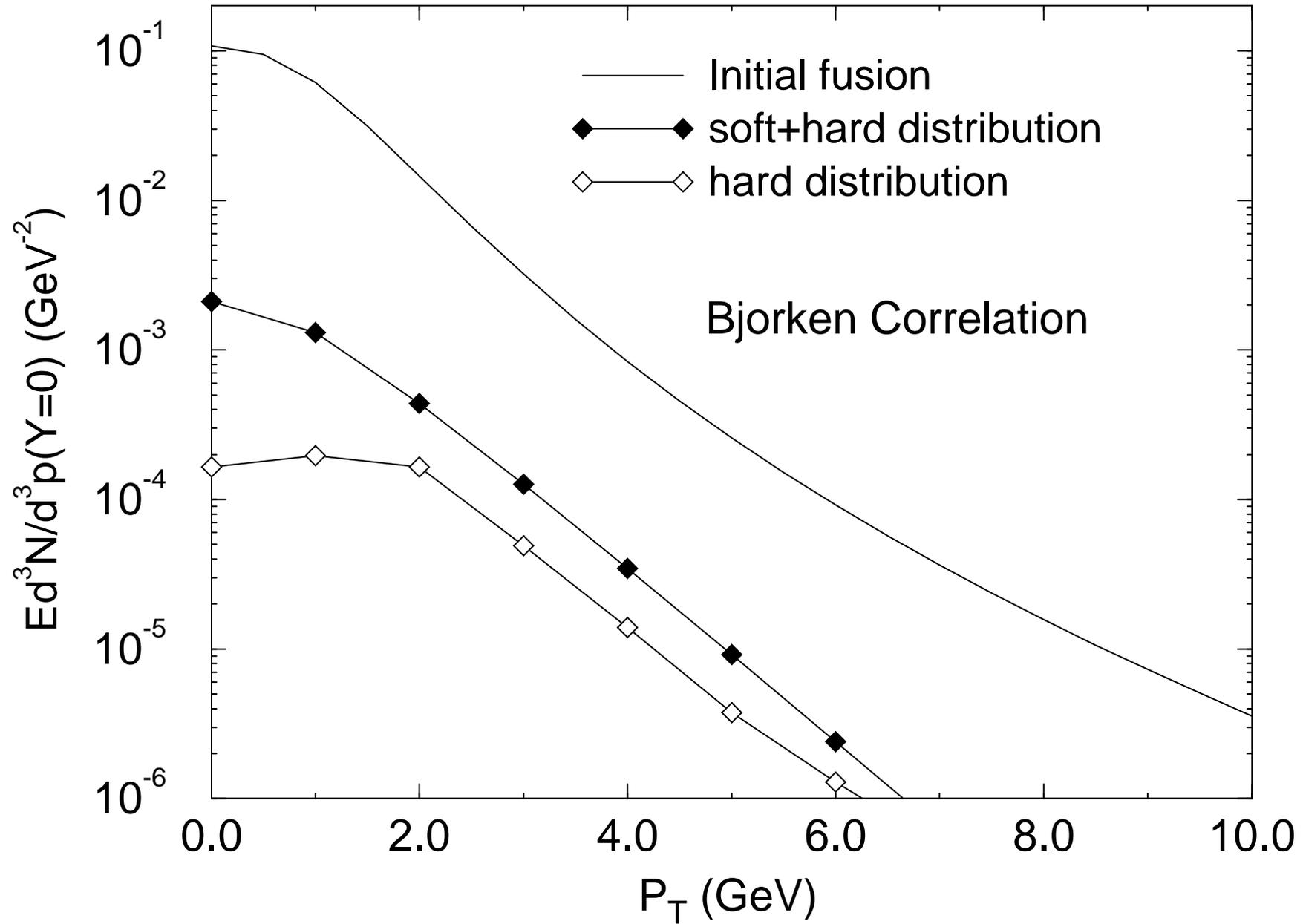
\item[Fig.6]
The distribution $\left( E d^3N/d^3p \right)_{y=0}$ of charm quark production
using $\delta(\eta-y)$-correlation is plotted as a function of $p_\perp$.  The
solid curve is the initial charm production.  The curve labelled with filled
diamonds is the pre-equilibrium contribution including both the  soft
($p_\perp<2$ GeV) and hard ($p_\perp> 2$ GeV) components of the mini-jet
gluons.  The curve labelled with unfilled diamonds is the pre-equilibrium
contribution including only the hard component.
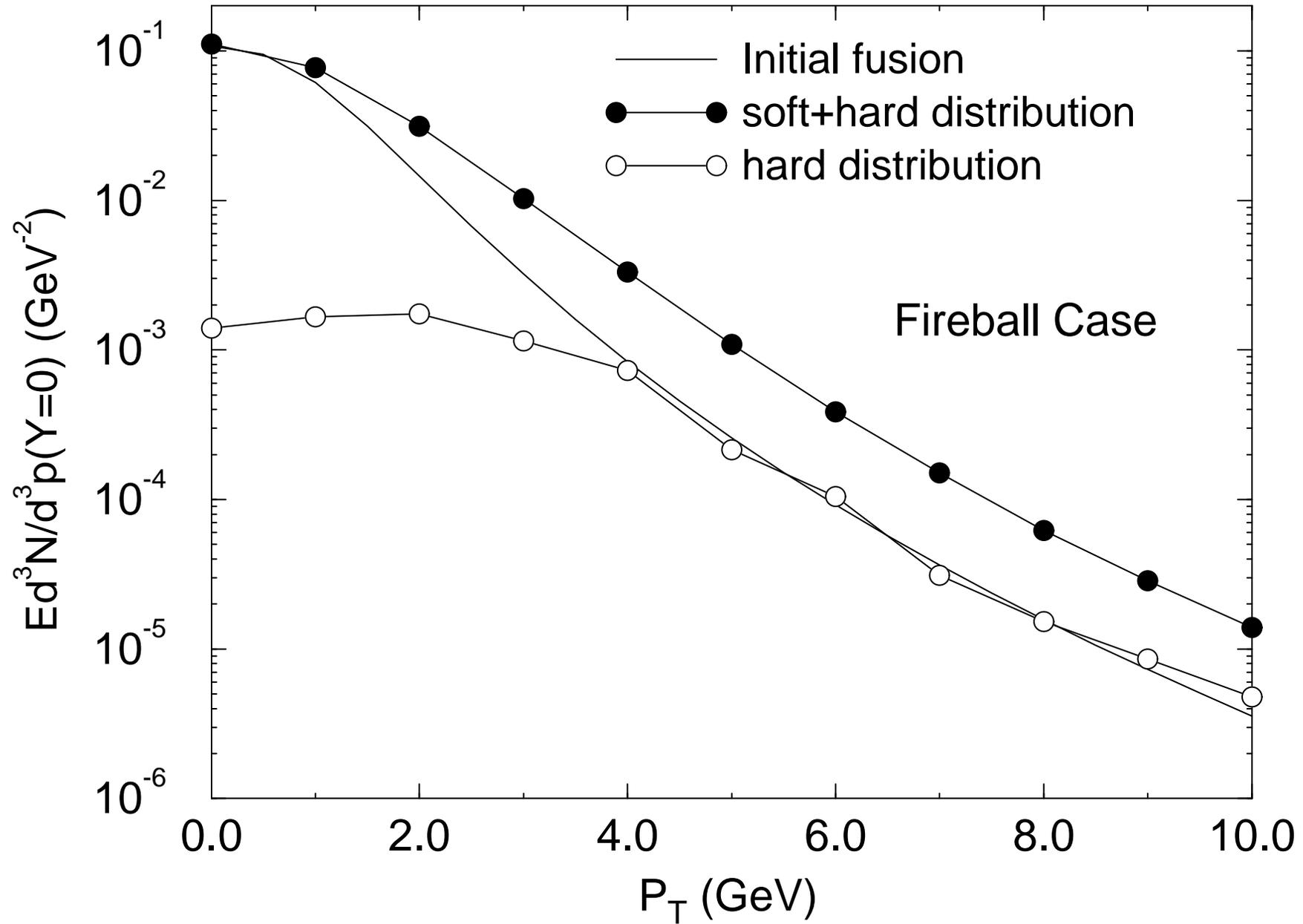
\item[Fig.7]
The distribution $\left( E d^3N/d^3p \right)_{y=0}$ of charm quark production
for the Uncorrelated case is plotted as a function of $p_\perp$.  The solid
curve is the initial charm production.  The curve labelled with filled
circles is the pre-equilibrium contribution including both the  soft
($p_\perp<2$ GeV) and hard ($p_\perp> 2$ GeV) components. The curve labelled
with unfilled circles is the pre-equilibrium contribution including only the
hard component.
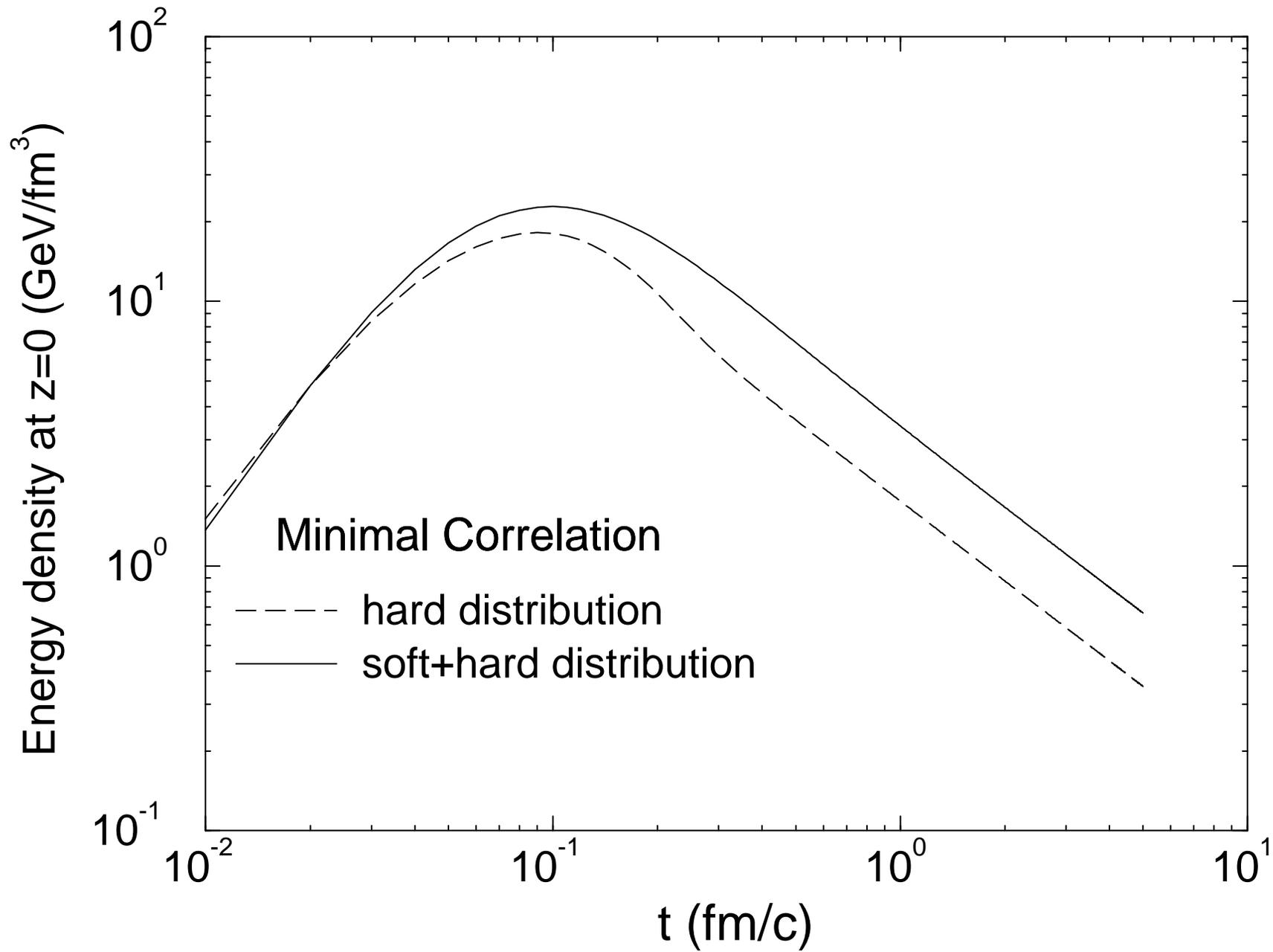
\item[Fig.8]
The energy density at $z=0$  is plotted as a function of proper time assuming
minimal correlations and Lorentzian formation probability. The solid curve
includes both soft and hard components while  the dashed curve is calculated
using the hard distribution and includes only the hard component.
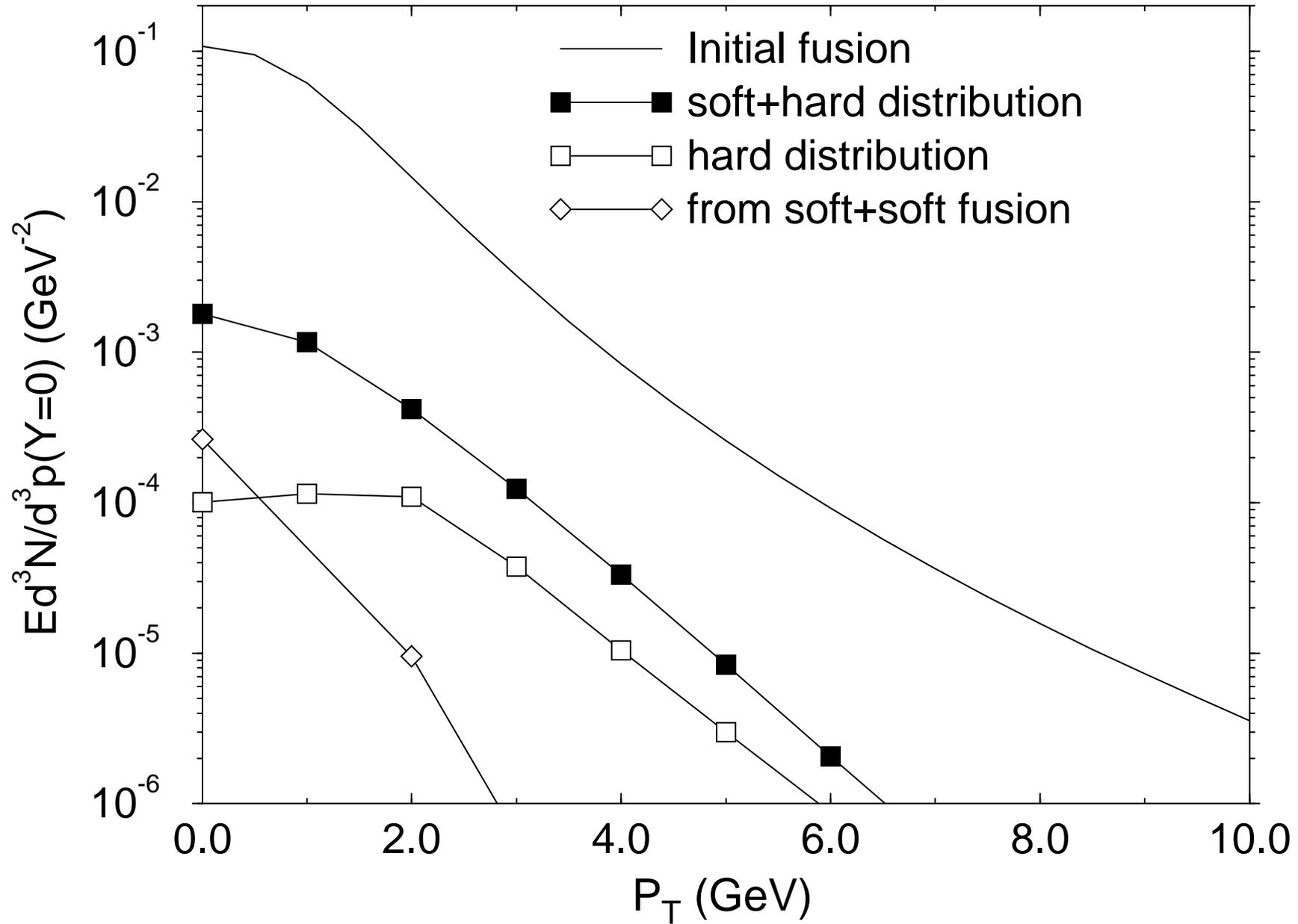
\item[Fig.9]
The distribution $\left( Ed^3N/d^3p \right)_{y=0}$ of charm quark production
using minimal $\eta-y$ correlations is plotted as a function of $p_\perp$. The
curve  labelled with filled squares include both components while that labelled
with unfilled squares include only the fusion of hard gluons. The curve
labelled with diamonds shows the contribution from fusion of soft gluons both
with  $p_\perp<2$GeV. This  shows that the pre-equilibrium contribution mainly
comes from the fusion of soft and hard gluons.
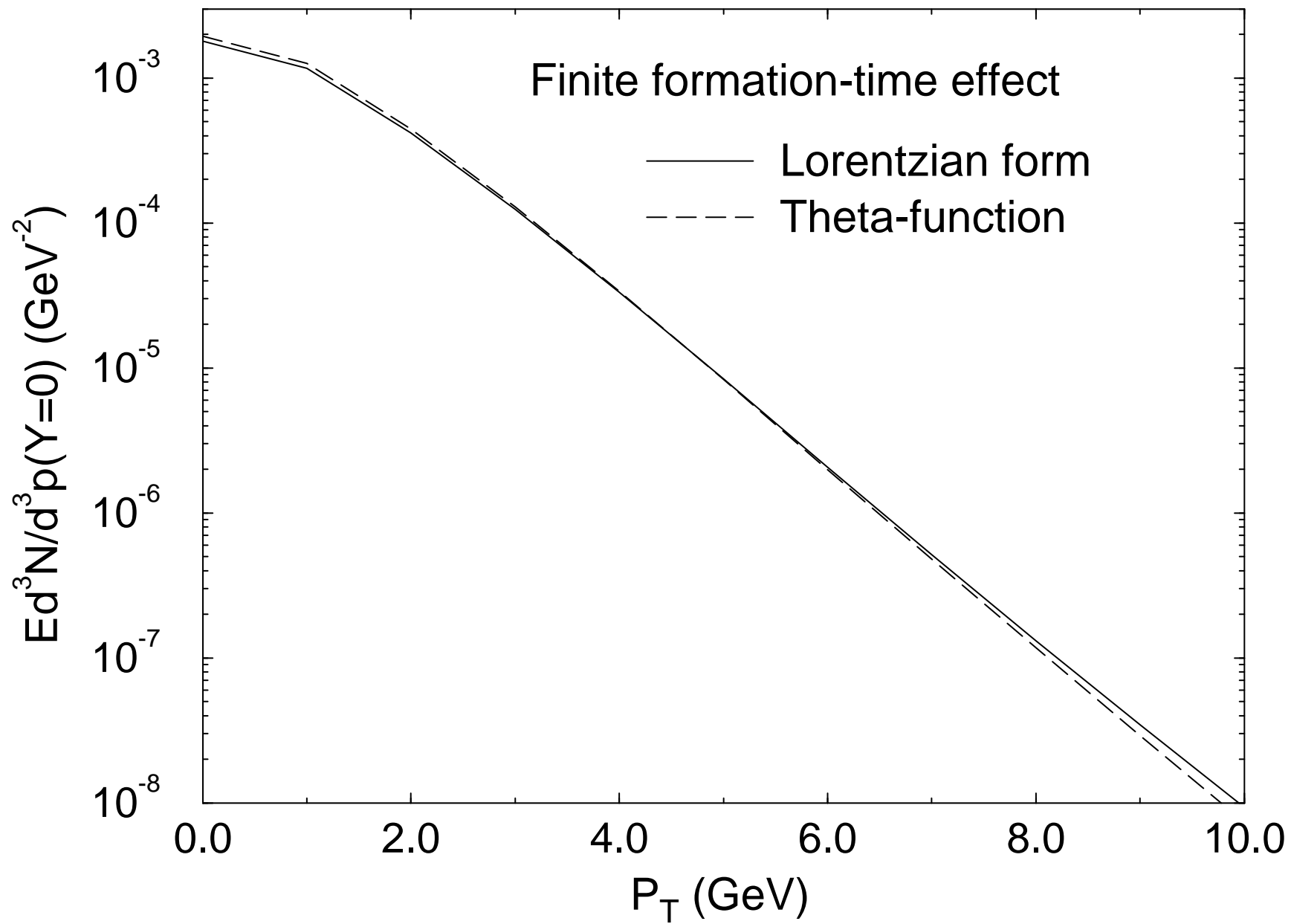
\item[Fig.10]
The distribution $\left( Ed^3N/d^3p \right)_{y=0}$ of charm quark production
using different formation-time probability distributions. The solid curve is
obtained using the Lorentzian form in equation~(\ref{EQ:lorentzian}), and the
dashed curve using the theta function form in equation~(\ref{EQ:theta}).
\end{description}
\end{document}